\documentclass[aps,prc]{revtex4}
\usepackage{epsfig}

\newcommand{\be}{\begin{equation}}
\newcommand{\ee}{\end{equation}}
\newcommand{\bea}{\begin{eqnarray}}
\newcommand{\eea}{\end{eqnarray}}
\newcommand{\bean}{\begin{eqnarray*}}
\newcommand{\eean}{\end{eqnarray*}}

\newcommand{\gapproxeq}{\lower
.7ex\hbox{$\;\stackrel{\textstyle >}{\sim}\;$}}
\newcommand{\lapproxeq}{\lower
.7ex\hbox{$\;\stackrel{\textstyle <}{\sim}\;$}}

\newcommand{\tenbar}{\mbox{$\mathrm{\overline{\bf 10}}$} }
\def\3bar{$\bar {\hbox{\bf 3}}$}

\begin{document}

\title{\bf Photoproduction of $\Theta^+$ and other pentaquark states}

\author{Frank E. Close$^1$\footnote{e-mail: F.Close1@physics.ox.ac.uk}
and Qiang Zhao$^2$\footnote{e-mail: Qiang.Zhao@surrey.ac.uk}
}

\affiliation{1) Department of Theoretical Physics,
University of Oxford, \\
Keble Rd., Oxford, OX1 3NP, United Kingdom}
\affiliation{2) Department of Physics,
University of Surrey, Guildford, GU2 7XH, United Kingdom}

\date{\today}

\begin{abstract}

We present constraints on the {\it relative} photoproduction 
cross sections of positive parity pentaquark states, 
$\Sigma_5$, $\Lambda_5$, and $N_5$,
based on a minimum phenomenology
gained in $\gamma N \to K \Theta^+$ and their baryon-meson 
couplings as in the work of Close and Dudek. 
The possibility of anomalous signals
in $\gamma p \to K^0_S (\Theta^+;\Sigma_d^+)$ is discussed. 
We emphasise the importance of comparing $\gamma N \to K \Theta^+$ 
with ``conventional" states such
as $\gamma N \to K \Sigma(1660)$.

\end{abstract}

\maketitle


\vskip 1.cm

Even though several photoproduction
experiments report signals
for the existence of the pentaquark $\Theta^+(1540)$
with $\sim 4-5 \sigma$
~\cite{spring-8,diana,clas,saphir,clas-2},
and narrow baryon states decaying into
$p K^0_S$ are also reported in other reactions~\cite{neutrino,hermes,svd,cosy},
the existence of the $\Theta^+(1540)$
is far from proved~\cite{dzierba,bes,fec}.
A definitive result
from a single experiment is still awaited.
Meanwhile there is no clear signal
of any other potential light pentaquark, save possibly the $\Xi^{--}$
states~\cite{na49},
though here again there are criticisms~\cite{fischer}.
In this Letter we propose a consistency check for the $\Theta^+$ and
related light pentaquark states
by focussing on the {\it relative} magnitudes of their photoproduction
cross sections.

Pentaquark models, designed to place the $\Theta^+$ $(q^4\bar{q})$ in a
\tenbar
of flavour, require
the $(qqqq)$ tetraquark to be in $\bar{{\bf 6}}_F$. When combined with
the
$\bar{{\bf 3}}_F$ of the $\bar{q}$
this implies pentaquark systems in both \tenbar$\oplus {\bf 8}$.
The narrow widths are then assumed to arise
from color-flavor-spin mismatch prior to falling apart
to form the octet meson and baryon~\cite{JW,maltman,cd}.

Close and Dudek proposed a scheme
to construct the flavor wavefunctions of pentaquark
states~\cite{fec,cd}.
This leads immediately to the relative strengths of all possible
channels to which a pentaquark state can couple.
We show here that some of these place severe constraints on the relative
production rates
for $\gamma N \to K\Theta^+; \ K\Sigma_5; \ K\Lambda_5$; and $\pi N_5$.
We suggest that dedicated experimental searches
may either reveal further pentaquarks or alternatively
highlight inconsistencies
that would call into question the interpretation 
of the narrow structure named 
$\Theta^+$ in $\gamma N \to K ``\Theta^+"$ as a pentaquark.

Several phenomenological studies of $\gamma N \to K \Theta^+$
have been proposed in the
literature~\cite{pr,hyodo,nam,liu-ko,oh,zhao-theta,za,n-t,y-c-j}.
Although these give quite different predictions in detail
due to their varied underlying assumptions,
they share certain essential common features
that enable robust constraints to be placed on the
{\it relative} production rates for various pentaquarks.
If the $\Theta^+$ should continue to be claimed in
photoproduction data, then these partner states
should also appear with similar rates. Failure to observe them would
require significant re-evaluation of the dynamics
of these hypothesised pentaquarks.

In this work, we will quantify the relative  cross sections for
photoproduction
of $\Theta$, $\Sigma_5$, $\Lambda_5$, and $N_5$.

The pentaquark wavefunctions in Ref.~\cite{cd} 
are written in a form that exposes 
their overlaps with the octet meson-baryon configurations to which the
pentaquark state will couple.
The explicit decomposition for $\Sigma_5$,
$\Lambda_5$, and $N_5$ is given in Table~\ref{tab-1}.
A tensor formalism that reproduces a subset of these results
(specifically
the decays
${\bf 8}_5 \to {\bf 8}_3 \otimes {\bf 8}_3$) has been given in
Ref.~\cite{oh2}.
As shown in Ref.~\cite{cd}, some selection rules emerge for their decay
branching ratios that may be used to help identify the
constitution of the states.
Future experiments should be able to examine such predictions
and provide tests of the underlying dynamics.
One advantage of the framework of Ref.~\cite{cd}
is that it provides a connection between the
pentaquark $\Theta^+$ and other pentaquark states such as
$\Sigma_5$ and $\Lambda_5$.
Therefore, information available for the $\Theta^+$ (even though still
limited) provides a guide for the production of
other pentaquark states.

To proceed, we will first summarize the phenomenology
for the $\Theta^+$ photoproduction in $\gamma n\to \Theta^+ K^-$
and $\gamma p\to \Theta^+ \bar{K^0}$.
Then we abstract the general features and evaluate their implications
for $\gamma N \to K (\Sigma_5;\Lambda_5)$, and $\pi N_5$.

For a positive parity $\Theta^+$ ($1/2^+$), a pseudovector effective
Lagrangian
is introduced for the $\Theta NK$ coupling~\cite{zhao-theta}:
\be
{\cal L}_{eff} = g_{\Theta NK}
\bar{\Theta} \gamma_\mu\gamma_5 \partial^\mu K N  + \mbox{h.c.},
\ee
where $N$ denotes the initial nucleon field, while
$\bar{\Theta}$ and $K$ denote the final state $\Theta^+$ and kaon.
We adopt $g_{\Theta NK}=1$ corresponding to total width
$\Gamma_{\Theta^+\to KN} \sim 1$ MeV in the calculations.
This choice is simply to set the scale; in all rates we
will show how they
rescale with a given total width $\Gamma_T$.

\begin{figure}
\begin{center}
\epsfig{file=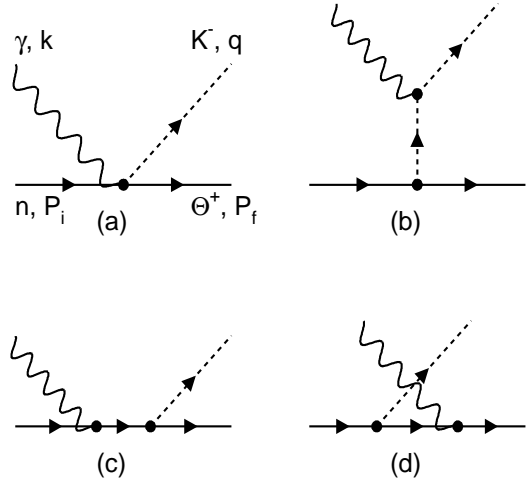, width=10cm,height=9.cm}
\caption{Feynman diagrams for pentaquark photoproduction in the Born
approximation. (a) and (b) will only contribute to the processes with
charged meson produced.
}
\protect\label{fig:(1)}
\end{center}
\end{figure}

The general assumption subsumed in
Refs.~\cite{pr,hyodo,nam,liu-ko,oh,zhao-theta,za,n-t,y-c-j} is that
for $\gamma n\to K^-\Theta^+$,
four transition amplitudes labelled by the Mandelstam variables
contribute in the Born approximation limit. These are
 shown by diagrams in Fig.~\ref{fig:(1)}: the contact term, {\it
t}-channel
kaon exchange, {\it s}-channel nucleon exchange and {\it u}-channel
$\Theta^+$ exchange. The differing output of the calculations in
Refs.~\cite{pr,hyodo,nam,liu-ko,oh,zhao-theta,za,n-t,y-c-j}
for magnitudes and energy dependences
are due to their various assumptions
on the magnitudes of, e.g., $g_{\Theta N K}$, $g_{\Theta N K^*}$, the role
of
the contact term and
importance of $s:u$ channels. However, some qualitative features can be
immediately understood, which are
sufficient for the generalisations in this present paper.

For $\gamma p\to \bar{K^0} \Theta^+$,
due to the absence of the contact and {\it t}-channel neutral-kaon
exchange,
the Born terms will only consist of
the {\it s}-channel nucleon exchange and {\it u}-channel $\Theta^+$
exchange. This feature leads to some model-independent predictions
for the Born contributions. In particular, due to the dominance
of the contact term in the Born limit,
cross sections for $\gamma n\to K^-\Theta^+$
will be significantly larger than those for $\gamma p\to \bar{K^0}
\Theta^+$.
The {\it t}-channel transition is also important over a wide
energy region.

In $K^0 \ (\bar{K^0})$ production,
the absence of the contact term and {\it t}-channel kaon
exchange requires
the possibility of $K^*$ exchange as essential. This is generally
assumed.
Therefore, we also include the $K^*$ exchange in this model
as the leading contribution in association with the Born terms.
The $K^*N\Theta$ interaction is given by
\be
{\cal L}_{\Theta N K^*}=g_{\Theta N K^*}\bar{\Theta}
(\gamma_\mu +\frac{\kappa_\theta^*}{2M_\Theta}
\sigma_{\mu\nu}\partial^\nu)V^\mu N   + \mbox{h.c.} \ ,
\ee
where $g_{\Theta N K^*}$ and $\kappa_\theta^*$ denote
the vector and ``anomalous moment" couplings, respectively.
The ``fall-apart" decay dynamics of Refs.~\cite{JW,maltman,cd}
implicitly
constrains these to be
$g^2_{\Theta N K^*} = 3 g^2_{\Theta N K}$ and $|\kappa_\theta^*|=0$.
Note that this will enhance the strength of the $K^*$ exchange
since most of the previous
literature~\cite{hyodo,nam,liu-ko,oh,n-t,y-c-j}
assume $|g_{\Theta N K^*}|/ |g_{\Theta N K}| \leq 1$.
The predicted cross sections also differ from
Refs.~\cite{zhao-theta,za},
where  $|\kappa^*_\theta|= |\kappa_\rho|=3.7$ was adopted,
based on an assumed similarity with $\rho NN$.
In this work, we adopt  $|\kappa_\theta^*|=0$, which is more
reasonable for
the physics of $(qqqq)\bar{q} \to N K/NK^*$ due to
the fall-apart dynamics.

The effective Lagrangian for the $K^*K\gamma$ vertex is given by
\be
{\cal L}_{K^*K\gamma}=\frac{ie_0g_{K^*K\gamma}}{M_K}
\epsilon_{\alpha\beta\gamma\delta}
\partial^\alpha A^\beta\partial^\gamma V^\delta K  + \mbox{h.c.} \ ,
\ee
where $V^\delta$ denotes the $K^*$ field.
For the M1 transition in the quark model one expects
that $g_{K^{*\pm}K^\pm\gamma} < g_{K^{*0}K^0\gamma}$,
the precise ratio being sensitively dependent
on the ratio $m_s/m_d$. Data agree with this;
the $\Gamma_{K^{*\pm}\to K^{\pm}\gamma}=50$ keV
implying $g_{K^{*\pm}K^\pm\gamma}=0.744$,
while $g_{K^{*0}K^0\gamma}=1.13$ is determined
by $\Gamma_{K^{*0}\to K^0\gamma}=117$ keV~\cite{pdg2000}.

Therefore, within the basic assumption that the meson-baryon-$B_5$
couplings (where $B_5$ denotes a generic pentaquark baryon)
are determined by ``fall-apart" topologies, we have constrained the
relative photoproduction cross sections for the various $B_5$ pentaquark
states.
The most obvious deficiency in the models is
the neglect of Reggeisation for the $K$-exchange. Thus, while the model
may be reliable near threshold, 
the energy dependence for $E_{\gamma}>3$ GeV will be
increasingly unreliable.
However, it is not essentially important for the purpose of this work.
Here, our interest is to make a connection between $\Theta^+$
and other photoproduced pentaquark states based on the available
information, in particular, the relative production rates
between the $\Theta^+$ and other pentaquark states.
Such relative rates, which are essentially controlled by the
``fall-apart" mechanism, should be less dependent on the
values of cross sections predicted by the phenomenology. Therefore, they
will provide a guide for the experimental
search for their signals.

Taking the above ingredients as input, we
classify production of pentaquark states $N_5$, $\Sigma_5$, and
$\Lambda_5$
into four classes:

i) charged meson production from neutron target:
\be
\label{n-c}
\left\{
\begin{array}{ccc}
\gamma n & \to & K^-\Theta^+ \nonumber\\
\gamma n & \to & K^+\Sigma_5^- \nonumber\\
\gamma n & \to & \pi^- p_5
\end{array}
\right. ;
\ee

ii) charged meson production from proton target:
\be
\label{p-c}
\left\{
\begin{array}{ccc}
\gamma p & \to & K^+\Lambda_5 \nonumber \\
\gamma p & \to & K^+ \Sigma_5^0 \nonumber\\
\gamma p & \to & \pi^+ n_5
\end{array}
\right. ;
\ee

iii) neutral meson production from neutron target:
\be
\label{n-n}
\left\{
\begin{array}{ccc}
\gamma n  &\to &  K^0\Lambda_5\nonumber\\
\gamma n  &\to &  K^0\Sigma_5^0\nonumber\\
\gamma n & \to & \pi^0 n_5 \nonumber\\
\gamma n  &\to & \eta_1 n_5({\bf 8}) \nonumber\\
\gamma n  &\to & \eta_8 n_5(\tenbar)
\end{array}
\right. ;
\ee

iv)  neutral meson production from proton target:
\be
\label{p-n}
\left\{
\begin{array}{ccc}
\gamma p & \to & \bar{K^0}\Theta^+\nonumber\\
\gamma p & \to & K^0 \Sigma^+_5 \nonumber\\
\gamma p & \to & \pi^0 p_5 \nonumber\\
\gamma p & \to & \eta_1 p_5({\bf 8}) \nonumber\\
\gamma p & \to & \eta_8 p_5(\tenbar)
\end{array}
\right. ;
\ee
where $\eta_1\equiv -(u\bar{u}+d\bar{d}+s\bar{s})/\sqrt{3}$
and $\eta_8\equiv (2s\bar{s}-u\bar{u}-d\bar{d})/\sqrt{6}$.

The following features can be abstracted for
the photoproduction of pentaquarks:

i) In the charged meson production reactions,
i.e. Eqs.~(\ref{n-c}) and (\ref{p-c}),
the cross section near threshold will be dominated by
the contact and {\it t}-channel charged-meson exchanges
in the Born approximation~\cite{zhao-theta}.

ii) In the neutral meson production reactions,
i.e. Eqs.~(\ref{n-n}) and (\ref{p-n}),
the contact and {\it t}-channel meson exchanges
will vanish since they are proportional to the outgoing meson charge.
Thus, only the {\it s}- and {\it u}-channel baryon-pole exchanges
will contribute to the Born terms.

iii) For the {\it s}-channel transitions, the difference
of the electromagnetic (EM)
couplings arises from different targets, i.e. proton or neutron.

iv) For the {\it u}-channel transitions, the difference of the EM
couplings
arises from the EM form factors of the pentaquark states,
theoretical estimates for some of which have been made
in Refs.~\cite{zhao-theta,h-d-c-z,l-h-d-c-z,cd2,kp,bijker}.
In photoproduction reactions, since the {\it u}-channel transitions
are generally suppressed due to the restricted kinematics,
the cross section is not a sensitive observable
to the pentaquark magnetic moments.
The {\it u}-channel contributions to the total cross section
is thus negligible, though they can produce
sizeable effects in polarization observables~\cite{zhao-theta,za}.

For $B_5 = \Theta^+,\Sigma_5,\Lambda_5$,
the {\it s} and {\it t}-channel amplitudes are proportional to $g_{B_5 N
K}$.
Therefore,
to the extent that the {\it u}-channel is negligibly small, the results
of Ref.~\cite{cd} 
imply that {\it the relative photoproduction cross sections 
for $\gamma N \to K B_5$ scale
in proportion to $g^2_{B_5 N K }$ and are correlated by effective
Clebsch-Gordan coefficients
apart from kinematic factors near threshold arising from mass
dependences}.

v) The {\it t}-channel vector-meson ($K^*$, $\rho$)
exchange can contribute to most of
those reactions. By comparing the charged and neutral meson
production processes, useful information about the
vector-meson-pentaquark
couplings can be gained.

\begin{figure}
\begin{center}
\epsfig{file=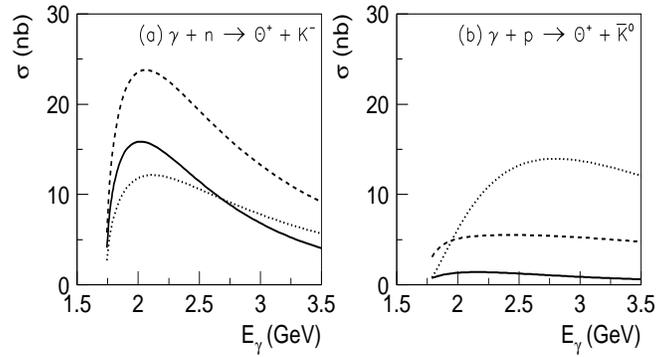, width=10cm,height=6.cm}
\caption{Total cross sections for $\Theta^+$ photoproduction in
$\gamma n\to \Theta^+ K^-$ and $\gamma p\to \Theta^+ \bar{K^0}$
with the total width of 1 MeV.
The solid curves are results in the Born limit
without $K^*$ exchanges, while the dashed and dotted curves are results
including $K^*$ exchanges of different signs.
For other total widths $\Gamma_T$ (in MeV),
the cross sections can be rescaled by $\Gamma_T/(1 \mbox{MeV})$. Thus,
e.g.,
if $\Gamma_T = 10$ MeV, the $\sigma$ will be 10 times larger.
}
\protect\label{fig:(2)}
\end{center}
\end{figure}

\begin{figure}
\begin{center}
\epsfig{file=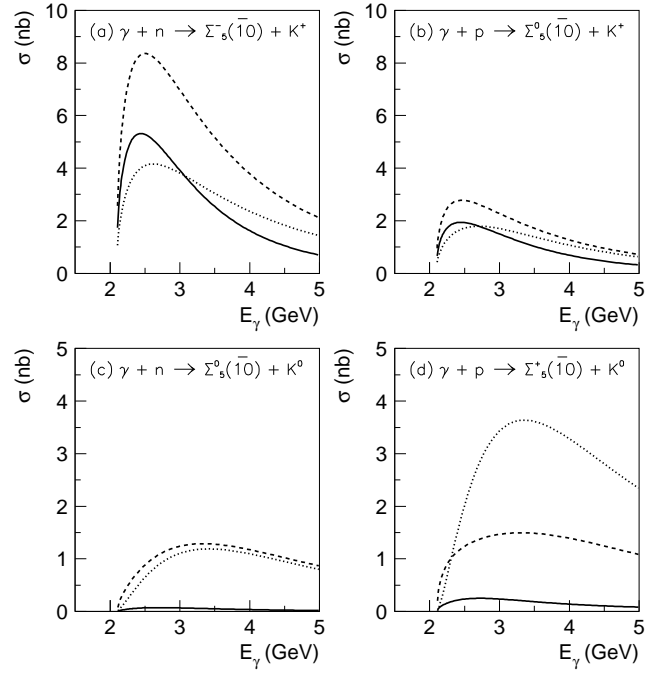, width=10cm,height=10.cm}
\caption{Total cross sections for $\Sigma_5(\tenbar)$
photoproduction in four reactions.
Notations are the same as Fig.~\ref{fig:(2)}.
}
\protect\label{fig:(3)}
\end{center}
\end{figure}

\begin{figure}
\begin{center}
\epsfig{file=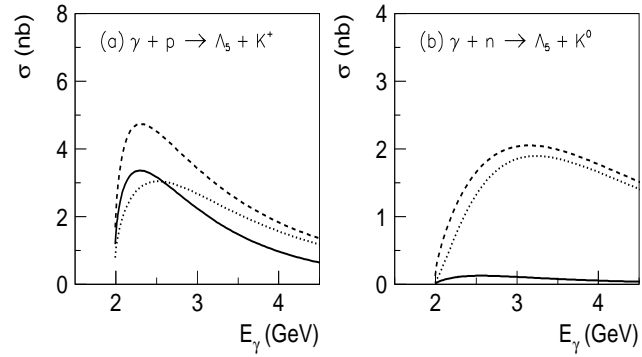, width=10cm,height=6.cm}
\caption{Total cross sections for $\Lambda_5$ photoproduction
on the proton and neutron. Notations are the same as Fig.~\ref{fig:(2)}.
}
\protect\label{fig:(4)}
\end{center}
\end{figure}

\begin{figure}
\begin{center}
\epsfig{file=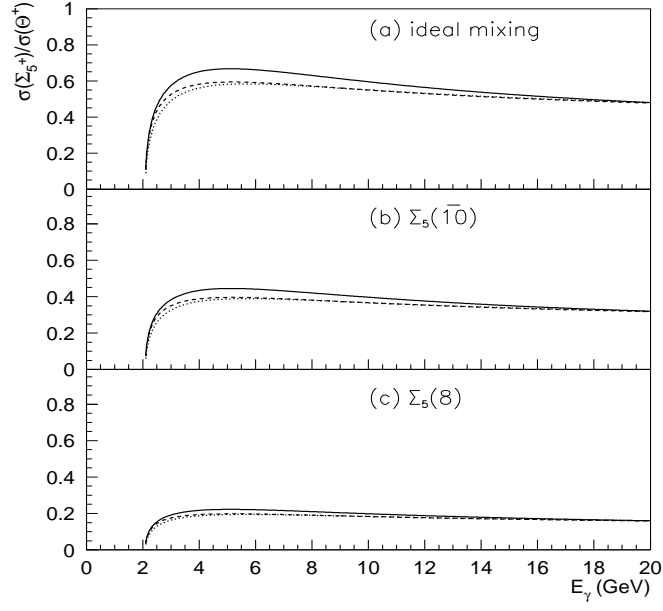, width=10cm,height=9.cm}
\caption{Energy dependence of the cross section ratios
for the ideally mixed $\Sigma^+_d$,  $\Sigma^+_5(\tenbar)$,
and $\Sigma^+_5({\bf 8})$. The Born terms are shown by the solid curves,
while the dashed and dotted curves denote the inclusion of $K^*$
exchange with $\pm$ signs.
}
\protect\label{fig:(5)}
\end{center}
\end{figure}

\begin{figure}
\begin{center}
\epsfig{file=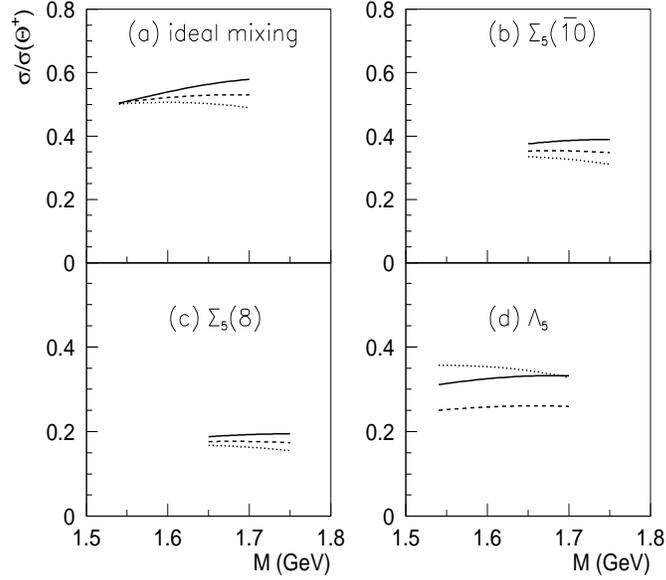, width=10cm,height=9.cm}
\caption{The cross section ratios
for the ideally mixed $\Sigma^+_d$,  $\Sigma^+_5(\tenbar)$,
$\Sigma^+_5({\bf 8})$, and $\Lambda_5$
as a function of the varying pentaquark masses
at $E_\gamma = 3$ GeV.
The Born terms are shown by the solid curves,
while the dashed and dotted curves denote the inclusion of $K^*$
exchange with $\pm$ signs.
}
\protect\label{fig:(6)}
\end{center}
\end{figure}

\begin{figure}
\begin{center}
\epsfig{file=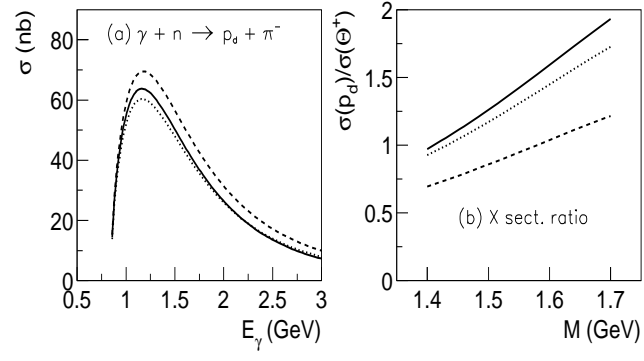, width=10cm,height=6.cm}
\caption{(a) Total cross sections for the nucleon pentaquark
$p_d$ photoproduction
with ideal mixing.
(b) The cross section ratios to the $\Theta^+$ as 
a function of the varying $p_d$ mass at $E_\gamma=3$ GeV.
The solid curves are results in the Born limit
without pion exchanges, while the dashed and dotted curves are results
including $\rho$ exchanges of different signs.
}
\protect\label{fig:(7)}
\end{center}
\end{figure}

The absence of Reggeisation means that the cross sections become
unreliable at high energy.
Thus the absolute cross sections
in Figs.~\ref{fig:(2)}, ~\ref{fig:(3)} and ~\ref{fig:(4)} are
only guides to show how existing models can be reproduced
and generalised. Our principal
results will be the {\it ratios} of production rates in
Figs.~\ref{fig:(5)}, Fig.~\ref{fig:(6)} and Fig.~\ref{fig:(7)}.

Concerning the strong couplings, we take the experimental
results for the $\Theta^+$ as a guide.
Since $\Theta^+$ can only strongly decay into $p K^0$ or $nK^+$,
the coupling constant $g_{\Theta NK}$ can be estimated by
$g_{\Theta NK}\equiv g_A M_n/f_\theta$, with
\be
\frac{f_\theta}{g_A}=\left(1-\frac{p_0}{E_n+M_n}\right)
\left[ \frac{|{\bf p}|^3(E_n+M_n)}
{4\pi M_\Theta\Gamma_{\Theta^+\to K^+ n}}\right]^{1/2},
\ee
where ${\bf p}$ and $p_0$ are the momentum and energy of the kaon
in the $\Theta^+$ rest frame, and $M_n$ and $E_n$ are the mass and
energy of the nucleon, respectively.

Within the fall-apart dynamics, $g_0\equiv g_{\Theta NK}$ is the
coupling strength setting the scale for
the strange-quark-associated pentaquark production and decay
amplitudes.
Thus, if there were no symmetry breaking in the masses, all pentaquark
states
considered here would have identical total widths. As $g^2_{B_5NK^*} =
3g^2_{B_5NK}$,
the overall ratios of cross sections survive even after the inclusion of
$K^*$ exchange contributions.

The partial width of decay channel $i$ is
\be
\Gamma_i=g_i^2 \frac{|{\bf p}|^3(E_f+M_f)}
{4\pi M_n^2 M_i}\left(1-\frac{p_0}{E_f+M_f}\right)^2 \ ,
\ee
where ${\bf p}$ and $p_0$ are momentum and energy of the final state
kaon
in the pentaquark rest frame; $E_f$ and $M_f$ are the energy
and mass of the final state baryon; $M_i$
is the pentaquark mass and $g_i$ is given by $g_0$ multiplied by the
relevant Clebsch-Gordan coefficient.
For instance, from Ref.~\cite{cd}
\be
\Theta^+(\tenbar) \to
\frac{1}{\sqrt 2}\{pK^0 -n K^+\} ,
\ee
which leads to
\be
\Gamma_T(\Theta^+)= \Gamma_{\Theta^+\to pK^0}
+\Gamma_{\Theta^+\to = nK^+}
\sim
1 \ \mbox{MeV} \ .
\ee

In Tables~\ref{tab-2} and ~\ref{tab-3}, the partial and isospin averaged
total widths are given, which are based on Ref.~\cite{cd}, but
modified by the phase space factors.

In the fall-apart dynamics all pentaquarks would have the same total
widths
in the absence of flavour symmetry breaking, which is illustrated
by Table~\ref{tab-1}.
For both $\Sigma_5(\tenbar)$ and $\Sigma_5({\bf 8}_5)$,
$M_{\Sigma_5}=1.7$ GeV is adopted before mixing.

For the production of $\Sigma_5^0$ the total width for the \tenbar
is about a factor of 2 bigger than that for the ${\bf 8}_5$.
The reason is that for channels, $\Xi^0 K^0$ and $\Xi^- K^+$, the decay
Clebsch-Gordans favor the
 ${\bf 8}_5$ relative to the $\tenbar$, but due to the
higher threshold of these two channels, the decay widths
are strongly suppressed.
This cuts into the width of the ``favoured" ${\bf 8}_5$ more than the
\tenbar,
thereby leading to the relatively larger width for the \tenbar.
The result is that after phase space is taken into account,
the total width of the $\Sigma_5(\tenbar)$ 
is about a factor of two larger
than that of the $\Sigma_5({\bf 8})$ and a factor of three
larger than the $\Theta^+$.

The mass of the $\Sigma_5$ could be an important issue in the search
for its signals. In Ref.~\cite{JW} mixing between the {\bf 8}$_5$ and
$\tenbar$ was assumed to be extreme such that the physical mass
eignestates are the light and heavy states with
``hidden $d\bar{d}$" and ``hidden $s\bar{s}$", respectively. We shall
label
these $\Sigma_d$ and $\Sigma_s$. If this is realised in practice there
is the
possibility that the $\Sigma_d$ has a mass similar to that of the
$\Theta^+$~\cite{JW}.
This could give interesting signals in $\gamma p \to K^0_S
(\Theta^+;\Sigma_d^+)$
arising from the primitive reactions
$\gamma p\to \bar{K^0}\Theta^+$ and $\gamma p\to K^0 \Sigma_5^+$.
In the case of degeneracy
the rate of $\Sigma_5^+$ production could be the same as $\Theta^+$
production in $\gamma p\to \bar{K^0}\Theta^+$
due to the interference between these two representations.
In general, for similar masses the $\Sigma_d$ production could be driven
by rescattering $\gamma p \to \bar{K^0}\Theta^+\to K^0 \Sigma_5^+$.
Such an ambiguity does not exist in the neutron target reaction, where
the production of $\Sigma_5^+$ is forbidden.
However, if the $\Sigma_5$ states are located significantly above the
$\Theta^+$, and/or rescattering is not important, 
as we shall show below, a kinematic suppression
can lead to a rather small rate for the production of $\Sigma_5$.

The production of $\Lambda_5({\bf 8})$ is also crucially
dependent on its mass. In this work, $M_{\Lambda_5}=1.65$ GeV
is adopted. The difficulty of isolating $\Lambda_5$
in either $\gamma p\to K^+\Lambda_5$ or $\gamma n\to K^0\Lambda_5$
is that the conventional octet states $\Lambda_3$ can be also produced
in
association with
$K^+$ or $K^0$. Indeed, a $P_{01}$ resonance $\Lambda(1600)$
with positive parity and a width about 150 MeV
is known to exist in this energy region and as such may mix with, and
hide,
the
primitive pentaquark state. However, in such a case the photoproduction
cross section for
the  $\Lambda(1600)$ would be at least as big as the pentaquark case
presented here since it
could then be produced through its pentaquark component as well as
``conventionally". Similar remarks apply
to $\Sigma^*$ states produced via any $\Sigma_5$ component.

To proceed, we will analyze reactions of
Eqs.~(\ref{n-c},~\ref{p-c},~\ref{n-n},~\ref{p-n})
based on our phenomenology for $\gamma p\to \bar{K^0}\Theta^+$
and $\gamma n\to K^-\Theta^+$.
As before we follow the fall-apart mechanism, where
$g^2_{\Sigma NK^*}=3g^2_{\Sigma NK}$,
and $g^2_{\Lambda NK^*}=3g^2_{\Lambda NK}$,
as a basic condition for the strange-quark-associated reactions.

In Fig.~\ref{fig:(2)}, total cross sections
for $\Theta^+$ with $\Gamma_T\sim 1 $ MeV are presented.
For other values of the total width, the estimates
can be rescaled by $\sigma \to \sigma\times\Gamma_T/(1\mbox{MeV})$.
As we repeatedly stress: the absence of Reggeisation means that the
energy dependence of the cross section in this and subsequent absolute
cross sections is unreliable for $E_{\gamma}>3$ GeV.

In the Born limit, the role played by the contact and {\it t}-channel
kaon exchange for the neutron target (i.e. $K^-$ production)
is highlighted by the significant difference between
the two solid curves.
Due to
$g^2_{\Theta NK^*}=3g^2_{\Theta NK}$~\cite{cd,maltman},
the contributions of the $K^*$ exchange are enhanced compared
with those in Ref.~\cite{zhao-theta}. In particular,
 $K^*$ exchange becomes dominant in $\gamma p\to \bar{K^0}\Theta^+$,
where the contact diagram and {\it t}-channel kaon exchange are absent.
The cross sections also show strong dependence on the
relative signs of the $K^*$ exchanges as illustrated by the
dashed and dotted curves.
Although simple $K$ and $K^*$ exchanges
in this model are constructive, the generalisation to
Regge trajectories leads to a non-trivial relative phase,
though the relative strength of three is
preserved due to its spin dependent origin.
Hence the cross section should
be considered as lying
within the range covered by the dashed and dotted curves.
This feature is consistent with other
model calculations.
If data continue to confirm the early suggestions
of comparable rates for the $\Theta^+$
production on the neutron~\cite{spring-8}
and proton target~\cite{saphir,clas-2},
this would be consistent with
a strong contribution from $K^*$ exchange.

In Fig.~\ref{fig:(3)}, total cross sections for
$\Sigma_5(\tenbar)$ in four reactions are presented.
The production of $\Sigma_5(\tenbar)$ is considered
with a mass of 1.7 GeV.
Cross sections for $\Sigma_5({\bf 8})$
production in general
are relatively suppressed by a factor of two compared with
those for $\Sigma_5(\tenbar)$.
The EM couplings for proton and neutron also result in
cross section differences. This feature is shown by comparing
Fig.~\ref{fig:(3)}(a) to~\ref{fig:(3)}(b), 
or comparing  Fig.~\ref{fig:(3)}(c) to~\ref{fig:(3)}(d).

These figures show that photoproduction cross sections for
$\Sigma_5(\tenbar)$
and $\Sigma_5({\bf 8})$
are much smaller than for $\Theta^+$ production. The suppression
comes from both the flavour coupling factors as shown in
Table~\ref{tab-1},
and kinematic effects in these low energy experiments;
e.g., the cross section difference between Fig.~\ref{fig:(2)}(b)
and Fig.~\ref{fig:(3)}(d) arises mainly from the different pentaquark
masses.
At higher energies, such as at HERMES~\cite{hermes}, only
the flavour coupling suppression is expected. So the relative
production rate for an ideal mixing $\Sigma_d$
between $\Sigma_5(\tenbar)$ and  $\Sigma_5({\bf 8})$, will lead to
$\sigma(\gamma n \to K^+ \Sigma^-_d )/\sigma(\gamma n \to K^-\Theta^+ )
\sim \sigma(\gamma p \to K^0 \Sigma^+_d )
/\sigma(\gamma p \to  \bar{K^0}\Theta^+)\sim 1/2$.

The analogous cross sections for photoproduction of $\Lambda_5$ with a
commonly assumed mass of 1.65 GeV are shown in Fig.~\ref{fig:(4)}.
The qualitative similarity and
quantitative difference between $\gamma p \to K^+\Sigma_5^0 $
[Fig.~\ref{fig:(3)}(b)] and
$\gamma p \to K^+\Lambda_5 $ [Fig.~\ref{fig:(4)}(a)]
is again due to the flavour coupling and kinematic effects.
If the spin-orbit partner $\Lambda_5$ ($J=3/2$) is below
1.9 GeV, and the fall-apart dynamics is correct, then this could
be another
narrow unmixed pentaquark~\cite{cd}, whose photoproduction cross section
should be similar or larger than those in Fig.~\ref{fig:(4)}.

Figures~\ref{fig:(2)},~\ref{fig:(3)} and~\ref{fig:(4)}
have shown the cross sections as a function of energy including
the Born terms and $K^*$ exchanges.
As repeatedly stressed, these are strongly model-dependent far above
threshold
due to the lack of Reggeisation.
However, visual inspection shows that the {\it ratios}
of $\Sigma_5$ and $\Lambda_5$ to $\Theta^+$ as a function of
$E_{\gamma}$ is much less sensitive to the $K^*$ exchanges
than the cross sections.
They are thus instructive for our purpose
of quantitizing the production rates of other pentaquark states
based on the limited knowledge about the $\Theta^+$.

In Fig.~\ref{fig:(5)}, the cross section ratio,
$\sigma(\gamma n \to K^+\Sigma^-_5)/\sigma(\gamma n \to  K^-\Theta^+)$,
for the production of $\Sigma_d$, $\Sigma_5(\tenbar)$,
and $\Sigma_5({\bf 8})$
to the $\Theta^+$ are presented.
The kinematic effects are shown by the energy evolution of the ratios.
If $\Sigma_5$ has the same mass as $\Theta^+$,
the ratios would be constants 1/2, 1/3, and 1/6 for
$\Sigma_d$, $\Sigma_5(\tenbar)$,
and $\Sigma_5({\bf 8})$, respectively, as discussed earlier.
Due to the mass difference, such constant ratios
occur at much higher energies as shown by the
convergence of the ratios for the Born terms and the charged $K^*$
exchanges.

The relative insensitivity of the production ratios of $\Sigma_5, \
\Lambda_5$
relative to the $\Theta^+$ is also useful for estimating 
the pentaquark production rates taking into account
the fact that their
masses have not been determined yet. Although ideal mixing gives
$\Sigma_d$ the same proportion of strange and nonstrange flavour masses
such that
naively one might expect $M_{\Sigma_d} = M_\Theta$,
a potential mass difference can arise due to the
$\bar{s}$ being isolated from the $(qqqq)$
in the $\Theta^+$ whereas it is in a diquark in the $\Sigma_d$.

We show in Fig.~\ref{fig:(6)}(a)-(c) the cross section ratios
of $\Sigma_d$, $\Sigma_5(\tenbar)$,
and $\Sigma_5({\bf 8})$
to $\Theta^+$ at $E_\gamma=3$ GeV (c.m. energy $W\simeq 2.55$ GeV)
as a function of their masses
within a theoretically predicted range:
  $\Sigma_5(\tenbar)$ and $\Sigma_5({\bf 8})$
are considered to have a mass range of 1.65 to 1.75 GeV;
an ideally mixed $\Sigma_d$ is taken to be within 1.54 to 1.7 GeV and
$\Sigma_s$ in 1.8 to 2 GeV.
Interestingly, it shows that in this energy region,
the cross section rates are not sensitive to
the pentaquark masses.
Hence, one should be cautious with interpreting 
any narrow baryon resonance
observed in $pK_S^0$ invariant mass in the absence of 
an accompanying strange hadron
that may constrain the strangeness of the narrow baryon. 
For example, in the HERMES data in particular~\cite{hermes},
we would expect the presence of $\Sigma^+$ accompanying
any $\Theta^+$.

The production of $\Sigma_s$ cannot be studied here since
it is now decoupled to the $NK$ channel.

The $\Lambda_5$ production ratio ($\gamma p\to K^+\Lambda_5 $)
to the $\Theta^+$ ($\gamma n\to  K^-\Theta^+$)
with a varying mass range 1.54 to 1.7 GeV is presented in
Fig.~\ref{fig:(6)} (d). It shows that the $K^*$
exchange could be the key process for the production of this state
if any sizeable signals are observed in experiment.

These ideas can be generalised to the cases of $N_5$
production with  $g_{N_5 N\pi(\rho)}=g_{\Theta N K(K^*)}$.
For the case of ideal mixing, the decomposition gives:
\be
\left\{
\begin{array}{ccl}
p_d & \to & -\frac{1}{6}\{ 3p\pi^0 -3\sqrt{2} n\pi^+ +\sqrt{6} p\eta_1
+\sqrt{3}p\eta_8\} \\
n_u & \to & \frac{1}{6}\{ -3\sqrt{2} p\pi^- +3 n\pi^0 +\sqrt{6} n\eta_1
+\sqrt{3} n\eta_8 \}
\end{array}\right. ,
\ee
for the proton and neutron pentaquarks. If we consider the non-strange
pentaquark $p_d$ and $n_u$ with mass assumed at 1.44 GeV~\cite{JW},
 the cross sections for $\gamma p \to n_u \pi^+$
and $\gamma n \to \pi^-p_d $ peak at around 70 nb
when $E_{\gamma} \sim 1.2$ GeV.
In Fig.~\ref{fig:(7)} (a), the
cross section for $\gamma n \to \pi^-p_d $ is shown.
The magnitude of cross sections is  larger than for $\Theta^+ K^-$
primarily because of the light $\pi$ mass.
Also, the cross section for $\gamma p \to \pi^+n_u $
is slightly larger than for the neutron reaction due to the presence
of the electric interaction.
Here, again the energy dependence of the cross sections
is strongly model-dependent due to the lack of
Reggeisation; but the ratios are still robust.
In Fig.~\ref{fig:(7)} (b), we show the relative
rates of $p_d$ to $\Theta^+$ for the mass range 1.4 to 1.7 GeV
at $E_\gamma=3$ GeV ($W\simeq 2.55$ GeV).

In summary: The work of Ref.~\cite{cd} provides a framework for
systematically
constructing pentaquark flavour wavefunctions with explicit
constraints on their relative phases and coupling strengths
in their strong decays into conventional octet baryons and mesons.
This then leads to a quantitative estimate of
other pentaquark photoproduction rates
based on the presently limited experimental and theoretical
information about the $\Theta^+$.
In particular,
continued observation of $\Theta^+$, and absence of $\Sigma_5, \ \Lambda_5$
at these rates would imply
that the pentaquark dynamics and masses are outside the currently
assumed range, or that these states do not exist.
Also, any experimental tests of the selection rules should shed
light on the pentaquark internal dynamics, in particular,
the fall-apart mechanism, which has been suggested to be important
for understanding the narrow widths.

Even if pentaquarks with non-exotic quantum numbers (such as
$\Lambda_5$, $\Sigma_5$ and $N_5$)
mix with conventional $qqq$ states, and as such are lost to a simple
spectroscopic search,
the photoproduction cross sections for these physical states should be
at least as big as those
calculated here, courtesy of any $B_5$ content. 
Thus, if photoproduction of states such as
$\Lambda(1600)$ or $\Sigma(1660)$  is not seen at
levels commensurate with Figs.~\ref{fig:(3)}, ~\ref{fig:(5)} and
~\ref{fig:(6)},
and no manifest pentaquark analogues are
seen either, then one would need to re-evaluate existing assumptions
about pentaquark spectroscopy and dynamics.

This work is supported,
in part, by grants from
the U.K. Particle Physics and
Astronomy Research Council, and the
EU-TMR program ``Eurodice'', HPRN-CT-2002-00311,
and the Engineering and Physical
Sciences Research Council (Grant No. GR/R78633/01).
FEC is indebted to A. Donnachie for discussions about Reggeised
photoproduction.


\begin{table}[ht]
\begin{tabular}{c|c|c}
\hline
Pentaquark & Representation & 
Decomposition into octet baryons and mesons\\[1ex]
\hline
$\Theta^+$ & $\tenbar$ & $\frac{1}{\sqrt 2}
\{ pK^0 -nK^+\}$ \\[1ex]
\hline
$\Sigma_5^+$ & $\tenbar $ & $\frac{1}{2\sqrt 3}
\{ \sqrt{3}\Sigma^+\eta_8 + \Sigma^+\pi^0 
-\Sigma^0\pi^+ -\sqrt{2} p\bar{K^0}
 +\sqrt{3}\Lambda\pi^+ +\sqrt{2}\Xi^0 K^+\} $\\[1ex]
 & ${\bf 8}_5$ & $\frac{1}{2\sqrt{6}}
\{\sqrt{6}\Sigma^+\eta_1 -\sqrt{3}\Sigma^+\eta_8
+ \Sigma^+\pi^0 -\Sigma^0\pi^+ -\sqrt{2} p\bar{K^0}
+\sqrt{3}\Lambda\pi^+ +2\sqrt{2}\Xi^0 K^+ \} $\\[1ex]
\hline
$\Sigma_5^0$ & $\tenbar $ & $-\frac{1}{2\sqrt{3}}
\{ \sqrt{3}\Sigma^0\eta_8 + n\bar{K^0} 
+pK^- -\sqrt{3}\Lambda\pi^0
+\Sigma^-\pi^+ -\Sigma^+\pi^- -\Xi^-K^+ -\Xi^0 K^0\}$ \\[1ex]
& ${\bf 8}_5$ & $-\frac{1}{2\sqrt{6}}
\{ -\sqrt{3}\Sigma^0\eta_8 +\sqrt{6}\Sigma^0\eta_1
+ n\bar{K^0} + pK^- -\sqrt{3}\Lambda\pi^0
+\Sigma^-\pi^+ -\Sigma^+\pi^- +2\Xi^-K^+ 
+2\Xi^0 K^0\}$ \\[1ex]
\hline
$\Lambda_5$ & ${\bf 8}_5$ &
$\frac{1}{2\sqrt{2}}\{ \Lambda\eta_8 + \sqrt{2}\Lambda\eta_1
+ n \bar{K^0} -pK^- +\Sigma^-\pi^+ +\Sigma^+\pi^- 
-\Sigma^0 \pi^0 \}$ \\[1ex]
\hline
$p_5$ & $\tenbar$ & $-\frac{1}{2\sqrt{3}}\{
\sqrt{3} p\eta_8 -\sqrt{2} n\pi^+ + p\pi^0 
+\sqrt{3}\Lambda K^+
+\sqrt{2} \Sigma^+ K^0 -\Sigma^0 K^+\}$ \\[1ex]
& ${\bf 8}_5$ & $-\frac{1}{2\sqrt{6}}\{
-\sqrt{6} p\eta_1 +2\sqrt{2} n\pi^+ 
- 2 p\pi^0 +\sqrt{3}\Lambda K^+
+\sqrt{2} \Sigma^+ K^0 -\Sigma^0 K^+\}$ \\[1ex]
\hline
$n_5$ & $\tenbar$ & $\frac{1}{2\sqrt{3}}\{
\sqrt{3}n\eta_8 + n\pi^0 -\sqrt{2} p\pi^- -\sqrt{3}\Lambda K^0
+\sqrt{2}\Sigma^-K^+ -\Sigma^0 K^0 \}$ \\[1ex]
& ${\bf 8}_5$ & $\frac{1}{2\sqrt{6}}\{
-\sqrt{6}n\eta_1 - 2 n\pi^0 + 2 \sqrt{2} p\pi^- 
-\sqrt{3}\Lambda K^0
+\sqrt{2}\Sigma^-K^+ -\Sigma^0 K^0 \}$ \\[1ex]
\hline
\end{tabular}
\caption{Decomposition of pentaquark flavour 
wavefunctions into octet baryons and mesons.}
\label{tab-1}
\end{table}

\begin{table}[ht]
\begin{tabular}{c|c|c|c}
\hline
Decay channels  & $\Theta^+$ & $\Sigma^+(\tenbar)$  &
$\Sigma^+({\bf 8}_5)$\\
\hline
$p K^0$ & 0.5 & * & * \\
\hline
$nK^+$ & 0.5 & * & * \\
\hline
$p \bar{K^0}$ & * & 0.57 & 0.28 \\
\hline
$\Sigma^+\pi^0$ & * & 0.47 & 0.23 \\
\hline
$\Sigma^0\pi^+$ & * & 0.47 & 0.23 \\
\hline
$\Sigma^+\eta_8$ & * & m.b.t. & m.b.t. \\
\hline
$\Lambda\pi^+$ & * & 1.74 & 0.87 \\
\hline
$\Xi^0 K^+$ & * & m.b.t & m.b.t. \\
\hline
$\Sigma^+\eta_1$ & * & * & m.b.t \\
\hline
Total (MeV) & 1.0 & 3.2 & 1.6 \\
\hline
\hline
\end{tabular}
\caption{Partial and isospin averaged total widths for $\Theta^+$
and $\Sigma_5^+$. Symbol ``*" stands for the decoupled channels, while
abbreviation ``m.b.t" for the case that the mass of pentaquark is below the
threshold of a decay channel. Total masses $M_{\Lambda_5}=1.65$ GeV and
$M_{\Sigma_5}=1.7$ GeV are adopted. We do not consider $\Sigma^* \pi$ as
this mode is forbidden by selection rules~\cite{cd}.
For the purpose of this table
we identify $\eta_8$ and $\eta_1$
with $\eta$ and $\eta^\prime$, respectively.}
\label{tab-2}
\end{table}

\begin{table}[ht]
\begin{tabular}{c|c|c|c}
\hline
Decay channels  & $\Lambda({\bf 8}_5)$ & $\Sigma^0(\tenbar)$
& $\Sigma^0({\bf 8}_5)$\\
\hline
$n \bar{K^0}$ & 0.32 & 0.28 & 0.14 \\
\hline
$p K^-$ & 0.32 & 0.28 & 0.14 \\
\hline
$\Lambda \eta_1$ & m.b.t & * & * \\
\hline
$\Lambda \eta_8$ & m.b.t & * & * \\
\hline
$\Lambda \pi^0$ & * & 1.75 & 0.87 \\
\hline
$\Sigma^0\pi^0$ & 0.54 & * & * \\
\hline
$\Sigma^-\pi^+$ & 0.54 & 0.46 & 0.23 \\
\hline
$\Sigma^+\pi^-$ & 0.54 & 0.46 & 0.23 \\
\hline
$\Sigma^0\eta_1$ & * & * & m.b.t \\
\hline
$\Sigma^0\eta_8$ & * & m.b.t. & m.b.t. \\
\hline
$\Xi^0 K^0$ & * & m.b.t & m.b.t \\
\hline
$\Xi^- K^+$ & * & m.b.t & m.b.t \\
\hline
Total (MeV) & 2.3 & 3.2 & 1.6 \\
\hline
\hline
\end{tabular}
\caption{Partial and isospin averaged total widths for $\Lambda_5$ and
$\Sigma_5^0$.
Notations are the same as Table~\ref{tab-2}.
The equality of
the widths for $\Sigma^0_5$ and $\Sigma^+_5$ 
is a numerical coincidence at these kinematics.}
\label{tab-3}
\end{table}




\end{document}